\pdfoutput=1
\documentclass[11pt,a4paper]{article}
\usepackage{amsmath,amssymb,amsthm,dsfont}
\usepackage[sc]{mathpazo}
\usepackage[margin=1in]{geometry}
\usepackage{cite}
\usepackage[colorlinks=true,allcolors=blue]{hyperref}

\newcommand{\X}{\mathcal{X}}
\newcommand{\Y}{\mathcal{Y}}
\newcommand{\id}{\mathds{1}}
\newcommand{\Tr}{\operatorname{Tr}}
\newcommand{\Sch}{\operatorname{Sch}}
\newcommand{\ket}[1]{|#1\rangle}
\newcommand{\bra}[1]{\langle #1|}
\newcommand{\ip}[2]{\langle #1|#2\rangle}

\newtheorem{theorem}{Theorem}
\newtheorem{corollary}{Corollary}
\theoremstyle{remark}
\newtheorem*{remark}{Remark}

\title{\Large\bf Simple proof of conditions requiring maximal entanglement for local state verification and discrimination}
\author{Somshubhro Bandyopadhyay\thanks{\texttt{som@jcbose.ac.in}}\\[1ex]
\small\it Department of Physical Sciences, Bose Institute, Kolkata 700091, India}
\date{}

\begin{document}
\maketitle

\begin{abstract}
For deterministic state verification by separable measurements, the Schmidt rank of an entangled resource must be at least that of the state being verified [N. Yu, R. Duan, and M. Ying, IEEE Trans. Inf. Theory \textbf{60}, 2069 (2014)], and if the two are equal, the resource must be maximally entangled [S. Akibue, J. Miyazaki, and H. Osaka, Lett. Math. Phys. \textbf{116}, 110 (2026)]. The latter was derived from a general theorem proved using algebraic geometry. In this paper, we give a simple, self-contained proof of both statements using only linear algebra. We use them to show that a generic orthogonal basis of $\mathbb C^d\otimes\mathbb C^d$ can be perfectly distinguished by LOCC using a pure resource in $\mathbb C^{d'}\otimes\mathbb C^{d'}$, with $d'\le d$, if and only if $d'=d$ and the resource is maximally entangled. The same holds for local implementation of a generic orthogonal measurement.
\end{abstract}

\section{Introduction}\label{sec:intro}

In quantum information theory, information is encoded in quantum states and extracted through measurements. In the distant laboratory paradigm, the setting of distributed quantum information processing, parts of a composite quantum system are shared between separated observers, and measurements are restricted to local operations and classical communication (LOCC) \cite{Chitambar2014}. In an LOCC protocol, the observers perform quantum operations on their local subsystems and communicate through classical channels, but cannot exchange quantum information. Thus not all measurements, e.g., entangled measurements, can be realized by LOCC. Such limitations can, however, be overcome using shared entanglement as a resource.

In this paper we study two fundamental quantum information processing tasks, state verification and state discrimination, within the framework of LOCC. In a state verification problem, the objective is to decide whether the shared system is in a given pure state $\ket{\varphi}$ or in a state orthogonal to it. If $\ket{\varphi}$ is entangled, then this cannot be accomplished with certainty \cite{Horodecki2003,YuDuanYing2014}. The reason is that the desired measurement is entangled and, therefore, cannot be implemented exactly by LOCC. In a state discrimination problem, the objective is to identify a state chosen from a known set of orthogonal states $\ket{\psi_1},\dots,\ket{\psi_N}$. While any two orthogonal pure states can be perfectly distinguished by LOCC \cite{Walgate2000}, the same does not hold in general for three or more states. For example, a complete orthogonal basis containing one or more entangled states is locally indistinguishable \cite{WalgateHardy2002,Horodecki2003}.\footnote{Locally indistinguishable states need not form a basis, nor be entangled. Examples are three Bell states \cite{Ghosh2001}, the states of an unextendible product basis \cite{Bennett1999UPB}, and the complete product basis of \cite{Bennett1999}.} Consequently, information encoded in such a basis is not completely accessible to spatially separated observers, even when they are allowed unrestricted classical communication.

Note that the two tasks are related. A protocol that perfectly distinguishes an orthogonal basis $\{\ket{\psi_i}\}$ also verifies each $\ket{\psi_j}$ by grouping the outcomes $i\neq j$. The converse need not hold. Both, however, can be accomplished with a shared maximally entangled state used as a resource. This follows from a simple teleportation protocol. One party teleports its share of the unknown state to the other \cite{Bennett1993}, who can then perform the appropriate measurement on both subsystems. Since entanglement is generally regarded as an expensive resource, a natural question is how much entanglement is necessary and sufficient for these tasks.

A lower bound on the entanglement needed for perfect local discrimination of an orthogonal basis of $\mathbb{C}^d\otimes\mathbb{C}^d$ is given by the average entanglement of the basis states \cite{Horodecki2003,BBKW2009} and is known as the entropy bound \cite{BRW2010}. For a basis of maximally entangled states the bound is tight and $\log_2 d$ ebits are necessary and sufficient \cite{Horodecki2003,BCJRWY2015,BR2024}. For $d=2$ improved lower bounds are known \cite{BBKW2009,BRW2010} but they stay close to the entropy bound. In general, the average entanglement of the basis states can take any value between $0$ and $\log_2 d$. One may therefore expect perfect local discrimination of a weakly entangled basis to require less entanglement.

Sometimes a single basis state determines the entanglement required of the resource. For example, if a two-qubit basis contains a maximally entangled state, perfect discrimination using a two-qubit resource requires it to be maximally entangled, even under positive-partial-transpose (PPT) measurements \cite{YuDuanYing2014}. In higher dimensions, if the basis contains at least one state of full Schmidt rank, perfect discrimination by one-way LOCC requires a resource of the same dimension to be maximally entangled \cite{BHN2018}, a result subsumed by \cite{AMO2026}.

Here we consider local state verification and discrimination using a pure entangled resource $\ket{\tau}$. Let $\Sch$ denote the Schmidt rank. The deterministic verification of a pure state $\ket{\varphi}$ by a separable measurement using $\ket{\tau}$ requires $\Sch(\tau)\ge\Sch(\varphi)$ \cite{YuDuanYing2014}, and if equality holds, $\ket{\tau}$ must be maximally entangled \cite{AMO2026}.\footnote{The latter result, proved recently, answers a question raised in \cite{YuDuanYing2014} for resources of the same dimension as the state.} We give a simple proof that establishes both results (Theorem~\ref{thm:AMO}). Unlike the algebraic-geometric argument of \cite{AMO2026}, our proof uses only linear algebra. In particular, the verification condition gives the Schmidt-rank inequality through a Schmidt-rank count, and when the Schmidt ranks are equal, measurement completeness leads to a trace identity that requires $\ket{\tau}$ to be maximally entangled. Similar results were proved earlier for the local implementation of nonlocal unitaries \cite{StahlkeGriffiths2011}.

We then apply Theorem~\ref{thm:AMO} to a state discrimination problem. We consider discriminating the elements of an orthogonal basis of $\mathbb{C}^d\otimes\mathbb{C}^d$ containing at least one state of Schmidt rank $d$ by separable measurements with the help of $\ket{\tau}\in\mathbb{C}^{d'}\otimes\mathbb{C}^{d'}$, where $d'\le d$. We show that perfect discrimination is possible if and only if $\ket{\tau}$ is a maximally entangled state of $\mathbb{C}^d\otimes\mathbb{C}^d$, and that the same holds for LOCC (Theorem~\ref{thm:main}).

Theorem~\ref{thm:main} also holds for almost every orthogonal basis of $\mathbb C^d\otimes\mathbb C^d$. This is because states of Schmidt rank less than $d$ form a lower-dimensional subset of measure zero, and therefore a Haar-random orthogonal basis (obtained by applying a Haar-random unitary to a fixed basis) consists entirely of full-Schmidt-rank states with probability one. The basis states may be arbitrarily weakly entangled, but what matters here is their Schmidt rank and not how entangled they are. Thus, with resources of local dimension at most $d$, extracting all the information encoded in a generic orthogonal basis of $\mathbb C^d\otimes\mathbb C^d$ by LOCC requires a maximally entangled resource. This answers the question of how much entanglement is necessary and sufficient for this task \cite{BBKW2009,BRW2010,BHN2018}.

\section{Preliminaries}\label{sec:setting}

Throughout this paper, the unknown state belongs to $\X_1\otimes\Y_1$ and the resource is a pure state $\ket{\tau}\in\X_2\otimes\Y_2$, where all spaces are finite-dimensional. Up to local unitaries,
\begin{equation}\label{eq:tau}
\ket{\tau}=\sum_\mu\lambda_\mu\ket{\mu\mu},\qquad \lambda_\mu\ge0,\qquad \sum_\mu\lambda_\mu^2=1.
\end{equation}
Let $\X=\X_1\otimes\X_2$ and $\Y=\Y_1\otimes\Y_2$, and let $W$ be the unitary operator mapping $(\X_1\otimes\Y_1)\otimes(\X_2\otimes\Y_2)$ to $\X\otimes\Y$, defined by
\begin{equation}\label{eq:W}
W\big(\ket{a_1b_1}\otimes\ket{a_2b_2}\big)=\ket{a_1a_2}\otimes\ket{b_1b_2}
\end{equation}
for $\ket{a_j}\in\X_j$ and $\ket{b_j}\in\Y_j$. The operator $W$ swaps the second and third subsystems. This formulation follows \cite{BCJRWY2015,BR2024}.

A measurement $\{P_1,\dots,P_N\}$ on $\X\otimes\Y$ is separable \cite{Bennett1999,Watrous2005} if each element has the form
\begin{equation}\label{eq:sep}
P_i=\sum_jA_{ij}\otimes B_{ij},
\end{equation}
where every $A_{ij}$ on $\X$ and every $B_{ij}$ on $\Y$ is positive semidefinite. Since a positive semidefinite operator is a positive combination of rank-one projectors, $A_{ij}$ and $B_{ij}$ can be taken to be of rank one.

\section{State verification}\label{sec:verif}

Let $\ket{\varphi}\in\X_1\otimes\Y_1$ be a pure state. In a state verification problem, the input is promised to be either $\ket{\varphi}$ or a pure state in its orthogonal complement. The verification test of $\ket{\varphi}$ is the projective measurement $\{\ket{\varphi}\bra{\varphi},\allowbreak\id-\ket{\varphi}\bra{\varphi}\}$.\footnote{In \cite{AMO2026} this test is called the optimal verification measurement, following \cite{Pallister2018}.} It is the only two-outcome measurement $\{M,\id-M\}$ on $\X_1\otimes\Y_1$ that satisfies
\begin{equation}\label{eq:verif}
\bra{\varphi}M\ket{\varphi}=1,\qquad
\bra{\chi}M\ket{\chi}=0\quad\text{for all }\ket{\chi}\perp\ket{\varphi}.
\end{equation}
The conditions $M\ge0$ and $\id-M\ge0$ imply $M\ket{\chi}=0$ and $M\ket{\varphi}=\ket{\varphi}$, respectively. Therefore, $M=\ket{\varphi}\bra{\varphi}$. It suffices to impose the second condition on a basis of the orthogonal complement of $\ket{\varphi}$. If $\ket{\varphi}$ is entangled, this test cannot be implemented by a separable measurement on $\X_1\otimes\Y_1$ alone, since $\ket{\varphi}\bra{\varphi}$ is not a separable operator.

We now suppose that $\ket{\varphi}$ is entangled and a resource $\ket{\tau}\in\X_2\otimes\Y_2$ is available. The input state is then $\sigma\otimes\ket{\tau}\bra{\tau}$, where $\sigma$ is a mixture of $\ket{\varphi}\bra{\varphi}$ and a state supported on the orthogonal complement of $\ket{\varphi}$.

Let $\{E,\id-E\}$ be a separable measurement on $\X\otimes\Y$ performed on $W(\sigma\otimes\ket{\tau}\bra{\tau})W^\dagger$. The probability of outcome $E$ is
\[
\Tr\!\left[EW(\sigma\otimes\ket{\tau}\bra{\tau})W^\dagger\right]=\Tr\!\left[M_E\sigma\right],
\]
where
\begin{equation}\label{eq:ME}
M_E=(\id\otimes\bra{\tau})W^\dagger EW(\id\otimes\ket{\tau})
\end{equation}
is the induced measurement element on $\X_1\otimes\Y_1$. Here $\id\otimes\bra{\tau}$ is the partial inner product with $\ket{\tau}$, defined by $\bra{x}(\id\otimes\bra{\tau})\ket{u}=(\bra{x}\otimes\bra{\tau})\ket{u}$ for all $\ket{x}\in\X_1\otimes\Y_1$. This measurement therefore implements the verification test of $\ket{\varphi}$ if $M_E$ satisfies \eqref{eq:verif}, i.e.\ if $M_E=\ket{\varphi}\bra{\varphi}$.

\begin{theorem}\label{thm:AMO}
If a separable measurement on $\X\otimes\Y$ implements the verification test of the pure state $\ket{\varphi}\in\X_1\otimes\Y_1$ using the resource $\ket{\tau}\in\X_2\otimes\Y_2$, then $\Sch(\tau)\ge\Sch(\varphi)$ \cite{YuDuanYing2014}. If equality holds, $\ket{\tau}$ is maximally entangled \cite{AMO2026}. The same holds for LOCC.
\end{theorem}

The proof below is direct and self-contained.

\begin{proof}
Suppose that a separable measurement $\{E,\id-E\}$ on $\X\otimes\Y$ implements the verification test of $\ket{\varphi}$, i.e.\ $M_E=\ket{\varphi}\bra{\varphi}$.

Let $E=\sum_k\ket{p_k}\bra{p_k}$ and $\id-E=\sum_k\ket{q_k}\bra{q_k}$, where $\ket{p_k}$ and $\ket{q_k}$ are product vectors (unnormalized) in $\X\otimes\Y$. We can therefore write $\ket{p_k}=\ket{\alpha_k}\otimes\ket{\beta_k}$ with $\ket{\alpha_k}\in\X_1\otimes\X_2$ and $\ket{\beta_k}\in\Y_1\otimes\Y_2$. Moreover, we can expand $\ket{\alpha_k}$ and $\ket{\beta_k}$ as
\[
\ket{\alpha_k}=\sum_\mu\ket{a_{k\mu}}\otimes\ket{\mu},\qquad
\ket{\beta_k}=\sum_\nu\ket{b_{k\nu}}\otimes\ket{\nu},
\]
with $\ket{a_{k\mu}}\in\X_1$ and $\ket{b_{k\nu}}\in\Y_1$, where $\{\ket{\mu}\}$ and $\{\ket{\nu}\}$ are the bases of $\X_2$ and $\Y_2$ in Eq.~\eqref{eq:tau}. Then
\begin{equation}\label{eq:key}
W^\dagger\ket{p_k}=\sum_{\mu\nu}\ket{a_{k\mu}b_{k\nu}}\otimes\ket{\mu\nu}\in(\X_1\otimes\Y_1)\otimes(\X_2\otimes\Y_2).
\end{equation}

\paragraph{Schmidt-rank inequality.}
The induced measurement element is given by Eq.~\eqref{eq:ME}. Substituting $E=\sum_k\ket{p_k}\bra{p_k}$ and using Eqs.~\eqref{eq:tau} and \eqref{eq:key}, we get $M_E=\sum_k\ket{v_k}\bra{v_k}$, where
\[
\ket{v_k}=\sum_\mu\lambda_\mu\ket{a_{k\mu}b_{k\mu}}.
\]
Since $\sum_k|\ip{\chi}{v_k}|^2=\bra{\chi}M_E\ket{\chi}=0$ for every $\ket{\chi}\perp\ket{\varphi}$, every $\ket{v_k}$ is proportional to $\ket{\varphi}$, i.e.
\begin{equation}\label{eq:contact}
\sum_\mu\lambda_\mu\ket{a_{k\mu}b_{k\mu}}=c_k\ket{\varphi},\qquad
\sum_k|c_k|^2=1.
\end{equation}
Since $\sum_k\ket{v_k}\bra{v_k}=\ket{\varphi}\bra{\varphi}$, the coefficients $c_k$ satisfy the normalization in Eq.~\eqref{eq:contact}.

This implies that at least one $c_k$ is nonzero. For each such $k$, Eq.~\eqref{eq:contact} expresses $\ket{\varphi}$ as a sum of at most $\Sch(\tau)$ product vectors. Therefore $\Sch(\tau)\ge\Sch(\varphi)$.

\paragraph{Necessity of maximal entanglement for equality.}
Suppose $\Sch(\varphi)=\Sch(\tau)=r$. Without loss of generality we restrict ourselves to the Schmidt supports. Accordingly, let $\X_1=\Y_1=\X_2=\Y_2=\mathbb C^r$.\footnote{Local projections of $E$ and $\id-E$ onto the Schmidt supports preserve separability, completeness, and the induced verification test.} Then $\ket{\tau}=\sum_{\mu=1}^r\lambda_\mu\ket{\mu\mu}$ with $\lambda_\mu>0$. Let $\ket{\Psi}=W(\ket{\varphi}\otimes\ket{\tau})$. From Eqs.~\eqref{eq:tau}, \eqref{eq:key}, and \eqref{eq:contact} we get
\[
\ip{\Psi}{p_k}=\big(\bra{\varphi}\otimes\bra{\tau}\big)W^\dagger\ket{p_k}=\sum_\mu\lambda_\mu\ip{\varphi}{a_{k\mu}b_{k\mu}}=c_k .
\]
Moreover, $\bra{\Psi}E\ket{\Psi}=\bra{\varphi}M_E\ket{\varphi}=1$, so
\[
\sum_k|\ip{q_k}{\Psi}|^2=\bra{\Psi}(\id-E)\ket{\Psi}=0 ,
\]
and every $\ip{q_k}{\Psi}$ vanishes.

Since $E+(\id-E)=\id$,
\begin{equation}\label{eq:expand}
\ket{\Psi}=\sum_k\ip{p_k}{\Psi}\ket{p_k}+\sum_k\ip{q_k}{\Psi}\ket{q_k}=\sum_{k:c_k\ne0}c_k^*\ket{p_k}.
\end{equation}
Therefore, by Eq.~\eqref{eq:key},
\[
\ket{\varphi}\otimes\ket{\tau}=W^\dagger\ket{\Psi}=\sum_{k:c_k\ne0} c_k^*W^\dagger\ket{p_k}=\sum_{k:c_k\ne0} c_k^*\sum_{\mu\nu}\ket{a_{k\mu}b_{k\nu}}\otimes\ket{\mu\nu}.
\]
Taking the partial inner product with $\ket{\mu\mu}\in\X_2\otimes\Y_2$ and using Eq.~\eqref{eq:tau}, we get
\begin{equation}\label{eq:mu}
\lambda_\mu\ket{\varphi}=\sum_{k:c_k\ne0}c_k^*\ket{a_{k\mu}b_{k\mu}}.
\end{equation}
Observe that Eqs.~\eqref{eq:contact} and \eqref{eq:mu} contain the same product vectors $\ket{a_{k\mu}b_{k\mu}}$. In Eq.~\eqref{eq:contact} they are summed over $\mu$ for fixed $k$, and in Eq.~\eqref{eq:mu} over $k$ for fixed $\mu$.

For $c_k\ne0$, Eq.~\eqref{eq:contact} expresses $\ket{\varphi}$ as a sum of $r$ product vectors $\ket{a_{k\mu}b_{k\mu}}$. We will now use Eq.~\eqref{eq:contact} to write each of these product vectors in terms of $\ket{\varphi}$ and substitute into Eq.~\eqref{eq:mu}.

Since $\Sch(\varphi)=r$, the vectors $\ket{a_{k1}},\dots,\ket{a_{kr}}$ form a basis of $\mathbb C^r$ with dual basis $\{\ket{\tilde a_{k\mu}}\}$ defined by $\ip{\tilde a_{k\mu}}{a_{k\nu}}=\delta_{\mu\nu}$. This allows us to apply the operator $\ket{a_{k\mu}}\bra{\tilde a_{k\mu}}\otimes\id$ to Eq.~\eqref{eq:contact}, which gives
\begin{equation}\label{eq:term}
\lambda_\mu\ket{a_{k\mu}b_{k\mu}}=c_k\big(\ket{a_{k\mu}}\bra{\tilde a_{k\mu}}\otimes\id\big)\ket{\varphi}.
\end{equation}
Multiplying Eq.~\eqref{eq:mu} by $\lambda_\mu$ and substituting Eq.~\eqref{eq:term} on the right-hand side, we get
\begin{equation}\label{eq:G}
(G_\mu\otimes\id)\ket{\varphi}=\lambda_\mu^2\ket{\varphi},
\end{equation}
where
\[
G_\mu=\sum_{k:c_k\ne0}|c_k|^2\ket{a_{k\mu}}\bra{\tilde a_{k\mu}}.
\]
Since $\ket{\varphi}$ has full Schmidt rank on $\mathbb C^r\otimes\mathbb C^r$, Eq.~\eqref{eq:G} implies $G_\mu=\lambda_\mu^2\id$. Taking the trace,
\begin{equation}\label{eq:trace}
r\lambda_\mu^2=\Tr G_\mu
=\sum_{k:c_k\ne0}|c_k|^2\ip{\tilde a_{k\mu}}{a_{k\mu}}=1.
\end{equation}
Thus every $\lambda_\mu^2=1/r$, proving that $\ket{\tau}$ is maximally entangled. The statement for LOCC follows since every LOCC measurement is separable.
\end{proof}

\begin{remark}
A maximally entangled resource of Schmidt rank $\Sch(\varphi)$ is also sufficient under LOCC. The input is first projected locally onto the Schmidt supports of $\ket{\varphi}$, and the task is then completed by teleportation.
\end{remark}

The proof also shows that the Schmidt-rank inequality follows only from the first outcome of the measurement, and that when the Schmidt ranks are equal, the necessity of maximal entanglement follows from the completeness condition.

Note that when $\Sch(\tau)>\Sch(\varphi)$, Eq.~\eqref{eq:contact} may contain more product vectors than the Schmidt rank of $\ket{\varphi}$. The vectors $\ket{a_{k\mu}}$ need not be linearly independent and the trace argument can no longer determine the Schmidt coefficients (see Sec.~\ref{sec:higher} for such an example).

\section{State discrimination}\label{sec:discr}

Now let $\X_1=\Y_1=\mathbb{C}^d$ and $\X_2=\Y_2=\mathbb{C}^{d'}$ with $d'\le d$, and let $\{\ket{\psi_1},\dots,\ket{\psi_{d^2}}\}$ be an orthogonal basis of $\X_1\otimes\Y_1$ containing at least one state of Schmidt rank $d$. The states to be distinguished are
\begin{equation}\label{eq:Psi}
\ket{\Psi_i}=W\big(\ket{\psi_i}\otimes\ket{\tau}\big)\in\X\otimes\Y,\qquad i=1,\dots,d^2.
\end{equation}
A measurement $\{P_1,\dots,P_{d^2}\}$ on $\X\otimes\Y$ perfectly distinguishes the states $\ket{\Psi_i}$ if
\begin{equation}\label{eq:perfect}
\bra{\Psi_j}P_i\ket{\Psi_j}=\delta_{ij}.
\end{equation}

\begin{theorem}\label{thm:main}
The states $\ket{\Psi_i}$ in \eqref{eq:Psi} can be perfectly distinguished by a separable measurement if and only if $\ket{\tau}$ is a maximally entangled state of $\mathbb C^d\otimes\mathbb C^d$. The same holds for LOCC.
\end{theorem}

\begin{proof}
Let $\ket{\psi_j}$ be an element of the basis with Schmidt rank $d$. Suppose a separable measurement $\{P_i\}$ perfectly distinguishes $\{\ket{\Psi_i}\}$. Then $\{P_j,\id-P_j\}$ is separable because $\id-P_j=\sum_{i\neq j}P_i$. By \eqref{eq:perfect}, outcome $j$ always occurs for $\ket{\Psi_j}$ and never occurs for any $\ket{\Psi_l}$ with $l\neq j$. Since the remaining $\ket{\psi_l}$ span the orthogonal complement of $\ket{\psi_j}$, this measurement implements the verification test of $\ket{\psi_j}$. Theorem~\ref{thm:AMO} gives
\[
d=\Sch(\psi_j)\le\Sch(\tau)\le d'\le d.
\]
Thus $d'=d$, and equality in Theorem~\ref{thm:AMO} requires $\ket{\tau}$ to be maximally entangled. This also proves necessity for LOCC. Sufficiency for LOCC, and therefore for separable measurements, follows from the teleportation protocol.
\end{proof}

Note that the only thing that matters in the proof of Theorem~\ref{thm:main} is a state with full Schmidt rank. The other states of the basis are not needed explicitly, except that they span its orthogonal complement.

Theorem~\ref{thm:main} has two corollaries. The first follows from the fact that almost every orthogonal basis of $\mathbb C^d\otimes\mathbb C^d$ consists of full-Schmidt-rank states.

\begin{corollary}\label{cor:generic}
Theorem~\ref{thm:main} holds for almost every orthogonal basis of $\mathbb C^d\otimes\mathbb C^d$.
\end{corollary}

Local discrimination of generic bases, not necessarily orthogonal, using shared entanglement was considered in \cite{LovitzJohnston2022}, where it was shown that a resource of Schmidt rank $d$ is necessary and sufficient for unambiguous discrimination.

The second corollary follows from the observation that a resource that enables perfect discrimination of a basis is also sufficient to implement a measurement in that basis in the sense of \cite{BBKW2009}, and vice versa.\footnote{In \cite{BBKW2009}, a protocol $P$ implements a measurement $M$ if it ``realizes $M$ exactly, in the sense that for any initial state of the system to be measured, $P$ yields classical outcomes with probabilities that agree with the probabilities given by $M$.'' It is not required that ``the measured system be collapsed into some particular state or even that it survive the measurement.'' For a complete orthogonal basis, a protocol that perfectly identifies every basis state necessarily has its basis projectors as the induced measurement elements.}

\begin{corollary}\label{cor:cost}
Let $\ket{\tau}$ be a pure resource of local dimension at most $d$. LOCC implementation of a measurement in almost every orthogonal basis of $\mathbb C^d\otimes\mathbb C^d$ using $\ket{\tau}$ requires $\mathcal{E}(\tau)=\log_2 d$ ebits, where $\mathcal{E}$ denotes the entanglement entropy.
\end{corollary}

For $d=2$, every entangled state has full Schmidt rank. So a two-qubit orthogonal basis either consists of product states or contains a state of full Schmidt rank. In the first case the basis is locally distinguishable \cite{WalgateHardy2002} and no resource is needed. In the second, a two-qubit resource must be maximally entangled for perfect local discrimination. This immediately applies to the two-qubit measurements considered earlier in \cite{BBKW2009,BRW2010} and to the elegant joint measurement \cite{Gisin2019}. For the latter, the need for a two-qubit maximally entangled resource was found numerically in \cite{AMO2026} and also follows from a general theorem there.

For $d\ge3$, the condition that the basis contain a state of full Schmidt rank cannot be relaxed. A basis may contain entangled states, none of which has full Schmidt rank, and still be perfectly distinguishable by LOCC with less than $\log_2 d$ ebits.\footnote{For example, the basis of $\mathbb C^3\otimes\mathbb C^3$ consisting of the Bell states \mbox{$(\ket{00}\pm\ket{11})/\sqrt2$}, \mbox{$(\ket{01}\pm\ket{10})/\sqrt2$}, and the product states $\ket{02},\ket{12},\ket{20},\ket{21},\ket{22}$ can be perfectly distinguished with one ebit.}

\section{Discussion}\label{sec:discussion}

\subsection{Resources of higher dimension}\label{sec:higher}

One may ask whether Theorem~\ref{thm:main} holds without the restriction $d'\le d$. Our proof requires this restriction because it relies on the equality case of Theorem~\ref{thm:AMO}. This may appear to be a limitation of our approach but the result, in fact, does not extend in general. The following examples show that resources in $\mathbb C^3\otimes\mathbb C^3$ with less than one ebit enable perfect discrimination of entangled orthogonal bases of $\mathbb C^2\otimes\mathbb C^2$ by separable measurements and by LOCC, respectively. These examples follow from constructions of \cite{StahlkeGriffiths2011} for implementing nonlocal unitaries.

Consider the orthogonal basis of $\mathbb C^2\otimes\mathbb C^2$:
\begin{equation}\label{eq:xy}
\ket{\psi_{xy}}=\tfrac12\big(\ket{00}+y\ket{01}+x\ket{10}+xy\,e^{-i\phi}\ket{11}\big),\qquad x,y=\pm.
\end{equation}
Now $\ket{\psi_{xy}}=U_\phi^\dagger\ket{xy}$, where $U_\phi=\operatorname{diag}(1,1,1,e^{i\phi})$. Therefore, applying $U_\phi$ followed by local measurements in the basis $\{\ket{+},\allowbreak\ket{-}\}$ distinguishes these states perfectly.

For $\phi=2\arccos(35/36)$, every state of this basis is entangled, with concurrence $|\sin(\phi/2)|\approx0.23$ \cite{Wootters1998}. The basis is therefore locally indistinguishable \cite{WalgateHardy2002,Horodecki2003}. A deterministic separable implementation of $U_\phi$ is given in \cite{StahlkeGriffiths2011} using the two-qutrit resource
\[
\ket{\tau}=\sqrt{0.81}\ket{00}+\sqrt{0.095}(\ket{11}+\ket{22}),
\]
where $\mathcal{E}(\tau)\approx0.89$ ebits.

For $\phi=0.08\pi$, the basis \eqref{eq:xy} is also entangled, with concurrence $\sin(0.04\pi)\approx0.13$ for each state. An LOCC protocol given in \cite{StahlkeGriffiths2011} implements $U_\phi$ using the two-qutrit resource
\[
\ket{\tau'}=\sqrt{0.8}\ket{00}+\sqrt{0.1}(\ket{11}+\ket{22}),
\]
where $\mathcal{E}(\tau')\approx0.92$ ebits. By Theorem~\ref{thm:main}, in both cases a two-qubit resource would have to be maximally entangled to enable perfect local discrimination.

\subsection{Open problems}

The higher-dimensional examples raise a natural question. When resources of higher dimension are allowed, how much entanglement is necessary and sufficient for perfect local discrimination of a given orthogonal basis? For a basis of maximally entangled states of $\mathbb C^d\otimes\mathbb C^d$, $\log_2 d$ ebits are necessary and sufficient. Beyond this little is known, even for specific types of bases. For example, one may first consider bases that contain just one maximally entangled state.

Another question concerns orthogonal bases of $\mathbb C^d\otimes\mathbb C^d$, $d\ge3$, where every basis state is entangled but none of them has full Schmidt rank. Suppose that $s<d$ is the largest Schmidt rank in the basis. Theorem~\ref{thm:AMO} implies that any resource enabling perfect discrimination must have Schmidt rank at least $s$, and a resource of Schmidt rank $s$ must be maximally entangled. The question is whether a maximally entangled resource of Schmidt rank $s$ is sufficient and, if not, what minimum Schmidt rank and entanglement are required.

A third question concerns incomplete sets of maximally entangled states. More than $d$ maximally entangled states of $\mathbb C^d\otimes\mathbb C^d$ cannot be perfectly distinguished by LOCC \cite{Nathanson2005}, and a complete basis of them requires $\log_2 d$ ebits \cite{Horodecki2003}. How much entanglement is needed to distinguish $N$ maximally entangled states with $d<N<d^2$? For three Bell states, a two-qubit resource must be maximally entangled \cite{BCJRWY2015}. Even for $d=2$ the question is open for resources of higher dimension, and for $d\ge3$ it is open in general.

\end{document}